\documentclass[conference]{IEEEtran}
\IEEEoverridecommandlockouts
\usepackage[usenames, dvipsnames]{color}
\usepackage[utf8]{inputenc}
\usepackage{multirow}
\usepackage{cite}
\usepackage{amsmath}
\usepackage{graphicx}
\usepackage{pgfplots}
\pgfplotsset{compat=newest}
\usetikzlibrary{pgfplots.statistics}
\usepackage{circuitikz}
\usetikzlibrary{circuits.logic.US, circuits.logic.IEC, positioning}
\pgfplotsset{compat=1.13}
\usepackage{verbatim}
\usepackage{cuted}
\usepackage{color}
\usepackage{environ}
\usepackage[font={footnotesize,sf}]{subcaption}
\usepackage{skull}
\usepackage{pgfplots}
\usepackage{tikz}
\usepackage[ruled,vlined,linesnumbered]{algorithm2e}

\begin{document}
\title{A Novel Approach for Fast and Accurate Mean Error Distance Computation in Approximate Adders}
\author{
\IEEEauthorblockN{Avishek Sinha Roy and Anindya Sundar Dhar}
\IEEEauthorblockA{Department of Electronics and Electrical Communication Engineering \\ Indian Institute of Technology, Kharagpur, WB 721302, India \\ sinharoyavishek@iitkgp.ac.in, asd@ece.iitkgp.ernet.in}
}
\date{\today}


\maketitle
\begin{abstract}
In error-tolerant applications, approximate adders
have been exploited extensively to achieve energy efficient system
designs. Mean error distance is one of the important error metrics
used as a performance measure of approximate adders. In this
work, a fast and efficient methodology is proposed to determine
the exact mean error distance in approximate lower significant
bit adders. A detailed description of the proposed algorithm
along with an example has been demonstrated in this paper.
Experimental analysis shows that the proposed method performs
better than existing Monte Carlo simulation approach both in
terms of accuracy and execution time.

\end{abstract}

\section{Introduction}
    In recent years, approximate computing has come forth as an encouraging solution to counter the rapid increase in energy consumption in modern-day applications like image processing and machine learning\cite{han2013approximate}. Approximate computing targeted for error-tolerant applications introduces selective approximation in arithmetic computations, which causes an occasional deviation from the theoretical output while achieving significant improvements in area, delay, and power. Adders being the key functional component in most of the error-tolerant applications have attracted a significant amount of research interest in this aspect\cite{1274006,zhu2010design,miao2012modeling,mahdiani2010bio,6241600,roy2016approximate}.
    
Approximate adders can be classified into two broad categories: Block-based approximate adders and approximate lower significant bit (LSB) adders. In block-based adders, the sum of each bit is calculated from speculative carry bits computed from previous LSB inputs\cite{li2014error}. The concept of block-based adders has originated from the principle that the probability of longer carry propagation is quite low for random input conditions\cite{1274006}. Examples of such type of adders are almost correct adder (ACA), error-tolerant adder, carry skip adder, carry speculative adder\cite{verma2008variable,zhu2009enhanced,kim2013energy,lin2015high}. On the other hand, in approximate LSB adders, the full adders in LSB positions are replaced with approximate adders. The full-adders in most significant bit positions are kept accurate. The approximate LSB adder includes lower-part OR adder (LOA) and approximate mirror adders (AMA) as examples\cite{mahdiani2010bio,6387646,miao2012modeling}. The block-based adders have high speed compared to approximate LSB adders. However, approximate LSB adders are comparably more hardware and power efficient\cite{jiang2015comparative}.

Various error-metrics have been introduced in literature to evaluate the efficiency of approximate adders along with traditional performance metrics such as area, delay and power\cite{liang2013new}. The error metrics include error rate (ER), mean error distance (MED), mean square error distance (MSED), mean relative error distance (MRED). For image processing applications, peak-signal-to-noise-ratio (PSNR) is typically used as a performance measure to evaluate image quality. It has been found that PSNR has higher dependence on error metric MED as compared to the ER\cite{liu2015analytical}. Though the approximate LSB adders have a very high ER compared to block based adders, the analysis presented in \cite{jiang2015comparative} shows that they have a moderate MED.

It is imperative to evaluate error statistics of various approximate adders for selecting an optimum design for a certain application. Several approximate circuit synthesis techniques have been proposed in literature where error metrics such as ER, MED are used as a constraining factor\cite{venkataramani2012salsa,wu2016efficient,vasicek2015evolutionary}. One of the biggest challenge is fast and accurate evaluation of error metrics. An exact MED calculation would require computation for all possible input combinations i.e $2^{2n}$ iterations for $n$-bit addition. For example, calculation of MED for a $12$-bit adder requires a simulation time of roughly 20 seconds running on Intel I5@3.2GHz processor core. Time-consuming exhaustive simulation can be avoided by adopting Monte Carlo sampling technique which provides near-exact measures of error metrics related to approximate adder designs\cite{venkatesan2011macaco,liu2015analytical}. Accurate methods to calculate error statistics in block based adders are presented in \cite{mazahir2017probabilistic} and \cite{DBLP:journals/corr/WuLGQ17}. However, to the best of our knowledge, no analysis has been presented in existing literature which shows accurate computation of error characteristics in  approximate LSB adders. In this article, we propose a novel accurate and efficient approach of MED calculation in approximate LSB adders.

The paper is structured as follows.The algorithm for MED computation in  approximate LSB adders is presented in Section II. An example for MED calculation of 2-bit approximate adder is also illustrated in this section. Analysis of the proposed algorithm in terms of iteration count and run-time is carried out in section III. Section IV summarizes the contribution of this paper. 
\section{Novel Fast and Accurate MED Computation Approach}
\begin{figure}[!ht]
        \centering
        \includegraphics[width=3.45in]{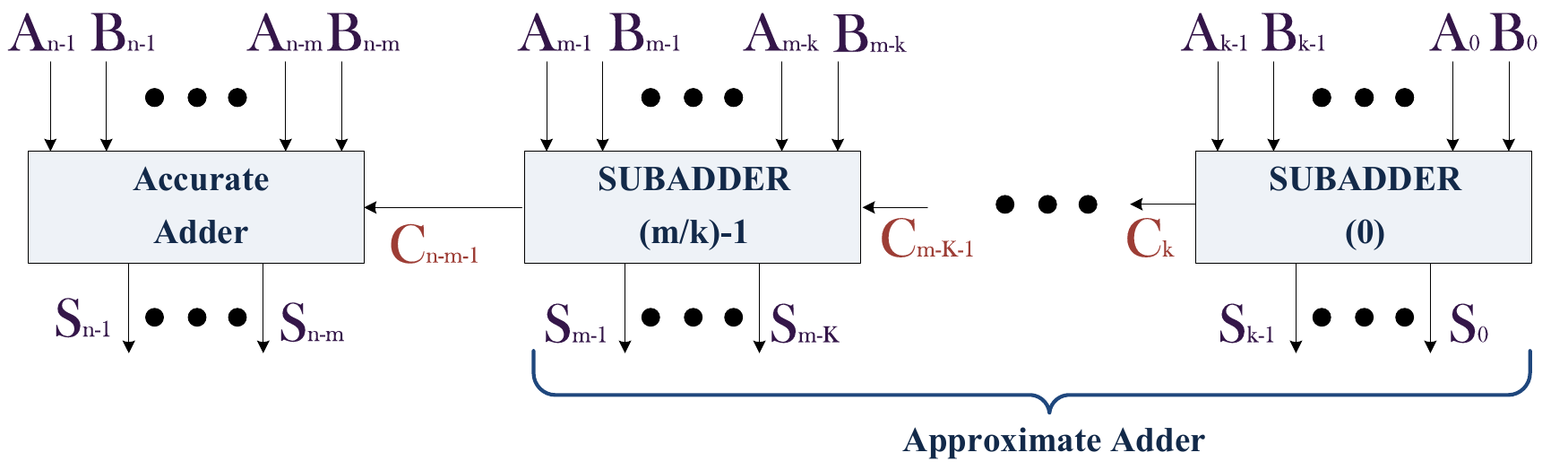}
        \caption{Generalized approximate LSB adder configuration}
        \label{fig:acc}
    \end{figure}

Error distance (ED) is defined as the absolute difference between the accurate computation result and the approximate result. The ED value averaged over all possible input combinations gives the mean error distance (MED) parameter. MED for $n$-bit adder is given by
\small{}
    \begin{equation}
\centering
\label{eq.mult}
\begin{split}
MED &= \frac{\sum_{i=1}^{2^{2n}}ED_{i}}{2^{2n}}
\\where ~ED &=  \left |  SUM_{ACC} -  SUM_{APP}\right |.
\end{split}
\end{equation}
\normalsize{}
ED is the absolute error distance whereas $SUM_{ACC}$ and $SUM_{APP}$ are the results of the accurate and the approximate addition of two $n$-bit inputs respectively. The generalized approximate LSB adder configuration considered for the MED computation is shown in Fig. \ref{fig:acc}. An $n$-bit approximate adder can be composed of $m$-bit approximate adders and $(n-m)$-bit accurate adder. Each $m$-bit approximate adder can be composed of several uniform $k$-bit approximate sub-adders with variable configurations. 
    	\vspace{-9pt}
\subsection{Proposed Algorithm MED\_Cal}
\begin{figure}[!ht]
        \centering
        \includegraphics[width=3in]{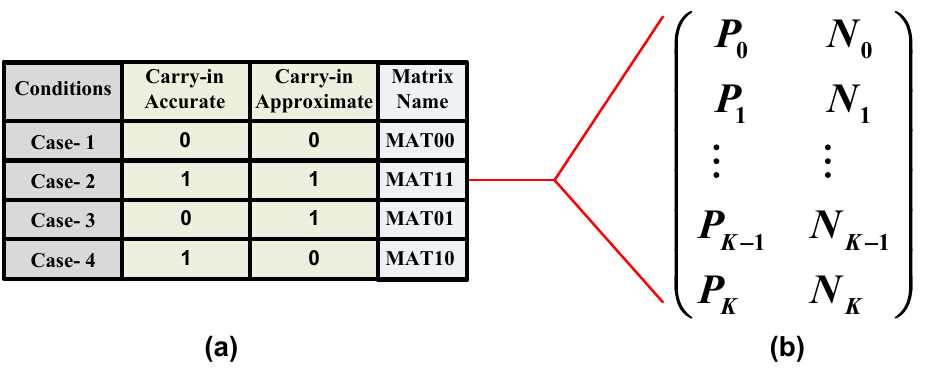}
        \caption{(a) Various carry-in conditions (b) Generalized matrix structure}
        \label{fig:cca}
    \end{figure}

\vspace{-2pt}
\begin{algorithm}[h]
\caption{\textit{\textbf{MED\_Cal}}}
\label{A34}
\footnotesize
\SetKwInOut{Input}{Input}
\SetKwInOut{Output}{Output}
\Input{$m$;$k$; $Truth\ table: Accurate\ Adder,Approximate\ Adder$.} 
\Output{$MED$.}
Initialize 4 matrices LMAT00 = $\left | 1 0 \right|$, LMAT01 = $\ \left | 0 0 \right|$, LMAT10 = $\ \left | 0 0 \right|$, LMAT11 =$\ \left | 0 0 \right|$.\\
Set LMAT00, LMAT01, LMAT10, LMAT11 as Input Matrices\\
\For{$i\ =\ 0\ to\ \frac{m}{k}-1$}{
\tcc{define p = $i\times$k; q = $(i+1)\times k -1$ }
$Initialize\ all\ elements\ of\ 4\ output\ matrices\ HMAT00,$\\ 
$HMAT01,HMAT10,\ HMAT11\ of\ size$ ($2^{q+1}\times2$) $to\ 0$\\
 \For{$all\ possible\ inputs$ $\left \{ A_{k-1:0}  B_{k-1:0}\right\}$}{
  \For{$all\ possible\ input\ carry\ conditions$ $\left \{ Cin_E  Cin_A\right\}$\\
  \tcc{$\left \{ Cin_E  Cin_A\right\}$ = \{00\} if i=0}}{
    $Compute\ CoutE,\ SumE_{q:p},\ CoutA,\ SumA_{q:p}.$\\
     $Compute\_Sum\_Difference()$;\\
     $Update\_Matrix\_Elements()$;\\
}
}
$Set\ HMAT00,\ HMAT01,\ HMAT10,\ HMAT11\ as\ Input$\\ $Matrices\ LMAT00,\ LMAT01,\ LMAT10,\ LMAT11$\\}
Calculate\_MED();
\end{algorithm}
	\vspace{-3pt}
The main objective of our proposed algorithm is to build a 2-D memory database of size $M\times2$. The parameter $M$ corresponds to the maximum absolute difference possible for an $n$-bit adder with $m$ LSB approximated which gives $M = 2^{m+1}$. Four such memory elements generically named as $MAT\left \{ CoutE\ CoutA \right\}$ as shown in Fig. \ref{fig:cca}a are created for four different carry-out conditions of $m^{th}$ bit. $CoutE$ and $CoutA$ represent exact and approximate carry-out bits respectively. Each of the row indices represent the differences between accurate and approximate addition for a certain combination of $n$-bit input $A$ and $B$. The elements in the matrix refer to the number of input combination of $A$ and $B$ that has the sum difference equal to the row index value. The first column specifies the positive difference while the second column indicates a negative difference in sum. The generalized memory structure of the final $MAT$ matrix is shown in Fig. \ref{fig:cca}b. The MED calculation approach proceeds from LSB to MSB bits of adder. We can consider the initial difference in sum between accurate and approximate adder as $0$ by assuming $m=-1$. Hence, four matrices of size ($2^{-1+1}\times2$) i.e ($1\times2$) are initialized. As we gradually approach towards higher significant bits from $0^{th}$ bit to $m^{th}$ bit, the matrix size also grows accordingly from ($1\times2$) to ($2^{m+1}\times2$).
    The proposed algorithm $MED\_Cal$ (Algorithm \ref{A34}) computes the MED of a generalized approximate LSB adder. The inputs to the algorithm are $m$ and $k$ which are defined earlier. The approximate and accurate adder functionality are also given as inputs in the form of a truth table. $LMAT$ and $HMAT$, which are matrices of type $MAT$ act as input and output matrices for each iteration of outermost \textbf{for} loop of $MED\_Cal$. All steps of the algorithm are illustrated below:
    \begin{itemize}
  \item \textbf{STEP-1\{\textit{Lines 1-2}\}}:Initially, four matrices $LMAT00$, $LMAT01$, $LMAT10$, $LMAT11$ are initialized to $\left | 1 0 \right|$, $\left | 0 0 \right|$, $\left | 0 0 \right|$  and $\left | 0 0 \right|$ respectively which would act as an input matrix for the $1^{st}$ iteration. Zero carry-in bit is considered for both accurate and approximate adder. This is the reason why the value stored at $(0,0)$ index of matrix $LMAT00$ is kept at $1$.
  \item \textbf{STEP-2\{\textit{Lines 3-5}\}}: In this step, four output matrices $HMAT00$, $HMAT01$, $HMAT10$, $HMAT11$  are defined, and all the elements of the matrices are initialized to zero. The size of the matrix is given as $2^{k(i+1)} \times 2$, where $i$ is the loop counter variable. The MSB and LSB of the approximate $k$-bit sub-adder considered in current iteration are represented by $q$ and $p$ respectively.
   \item \textbf{STEP-3\{\textit{Lines 6-9}\}}: Next, all possible input combinations of $k$-bit adder are generated. For each input combination, the output of accurate and approximate adder is computed for $4$ different carry-in conditions. The various outputs that are determined in this step includes accurate sum ($SumE_{k-1:0}$), approximate sum ($SumA_{k-1:0}$), accurate carry ($CoutA$), and approximate carry($CoutE$).
    \item \textbf{STEP-4\{\textit{Line 10}\}}: The  $Compute\_Sum\_Difference()$ function then evaluates differences in the accurate and approximate sum using Eqn.\ref{eq.diff1}.
    \begin{equation}
\centering
\label{eq.diff1}
\begin{split}
diff &= \sum_{j=p}^{q}2^{j}(SumE_j-SumA_j)
\end{split}
\end{equation}
 \item \textbf{STEP-5\{\textit{Lines 11-12}\}}: After $diff$ is computed, the function $Update\_Matrix\_Elements()$ updates $HMAT$ matrices by adding elements from input $LMAT$ matrices in each iteration. The input matrix is given by $LMAT\ \left \{ Cin_E Cin_A\right\}$ where $\{CinE,CinA\}$ corresponds to one of the four possible input carry conditions \{00;01;10;11\}. On the other hand, the output matrix is identified as $HMAT\ \{CoutECoutA\}$ where $CoutE$ and $CoutA$ are the carry out bits generated from respective $k$-bit accurate and approximate adder. An index mapping from input to output matrix has to be performed in this step before the contents of the LMAT matrix are added to HMAT matrix. Eqn. \ref{eq.im} illustrates the index mapping operation. The rows and column indices of LMAT matrix are represented by $r$ and $c$ where $r\in[0,2^p]$ and $c\in[0,1]$. After completion of each iteration, the $HMAT$ matrices are set as new input matrices $LMAT$. 
 
     \vspace{-8pt}
	\begin{equation}
\centering
\label{eq.im}
\scalebox{0.74}{$
	\begin{aligned}
\begin{tabular}{ll}
$LMAT(r,c) \rightarrow HMAT(r,c)$ &  $\,$ $if~diff = 0$  \\
$LMAT(r,0) \rightarrow HMAT(diff+r,0)$ & \multirow{2}{*}{\Big\} $if~diff > 0$} \\
$LMAT(r,1) \rightarrow HMAT(diff-r,0)$ &  \\
$LMAT(r,0) \rightarrow HMAT(\left |diff + r  \right |,1)$ & \multirow{2}{*}{\Big\} $if~diff < 0$} \\
$LMAT(r,1) \rightarrow HMAT(\left |diff - r  \right |,1)$ & 
\end{tabular}
	\end{aligned}$}
\end{equation}
\item \textbf{STEP-6\{\textit{Line 13}\}}: The steps $2-5$ are then repeated $\frac{m}{k}$ times until we get 4 matrix each of size $M\times2$. The error distance for the final matrix $HMAT \{CoutECoutA\}$ is then evaluated using function $Calculate\_MED()$. The cumulative ED for any matrix of type shown in Fig. \ref{fig:cca}b can be calculated using Eqn. \ref{eq.edcalc}. The parameter $\mu$ in this equation equals to $+1$ and $-1$ for unsigned number and signed numbers respectively. If $ED_{tot00}$, $ED_{tot01}$, $ED_{tot10}$, $ED_{tot11}$ corresponds to the cumulative ED computed from $HMAT00$, $HMAT00$, $HMAT00$, $HMAT00$ respectively, the final MED then can be computed using Eqn. \ref{eq.medcalc}.
\end{itemize}
\begin{equation}
\centering
\label{eq.edcalc}
\tiny{
ED_{tot} = \begin{cases}

((P_{_0} + N_{_0})\times0)+((P_{_1}+N_{_1})\times1)+\ldots \\
\qquad +((P_{_{M-1}}+N_{_{M-1}})\times (M-1)) \quad \textit{ if } CoutE=CoutA \\
\\
|\mu2^{^{m}}(CoutE-CoutA)(P_{_0} + \ldots + P_{_{M-1}} + N_{_0} + \ldots \\
\qquad  + N_{_{M-1}}) + ((P_{_0}-N_{_0})\times0) + ((P_{_1} - N_{_1})\times1) + \ldots \\ 
\qquad + ((P_{_{M-1}}-N_{_{M-1}})\times (M-1))| \quad \textit{ if } CoutE\neq CoutA 
\end{cases}}
\end{equation}
\vspace{5pt}
\scriptsize
    \begin{equation}
\centering
\label{eq.medcalc}
\begin{split}
MED &= \frac{ED_{tot00}+ED_{tot01}+ED_{tot10}+ED_{tot11}}{2^{2m}}
\end{split}
\end{equation}
\normalsize

	For a clear understanding of the algorithm, an example is presented for MED computation of an unsigned approximate adder with $m=2$, $k=1$ and $C_{in} = 0$. This assumption leads to a $2$-bit approximate adder with an LSB half-adder and an MSB full-adder. The truth-table of LSB and MSB adder of a random 2-bit approximate adder example are illustrated in Table. \ref{fig:app2bit}. Since initial carry-in is fixed to zero, there will be only one carry-in condition resulting in 4 iterations for $i=0$. For $i=1$, there will be $16$ iterations, $4$ iteration for all possible $4$ carry-in conditions. All the iterative steps are presented in Table \ref{aaex}.

\begin{table}[!ht]
\centering
\caption{Truth table: 2-bit approximate adder example}
\label{fig:app2bit}
\resizebox{.48\textwidth}{!}{
\begin{tabular}{ccccl|l|l|l|l|l|cc}
\cline{6-10}
\multicolumn{1}{l}{} & \multicolumn{1}{l}{} & \multicolumn{1}{l}{} & \multicolumn{1}{l}{} &  & $\mathbf{C_{in}}$ & $\mathbf{A_1}$ & $\mathbf{B_{1}}$ & $\mathbf{C_{out}}$ & $\mathbf{S_{1}}$ &  &  \\ \cline{6-10}
\multicolumn{1}{l}{} & \multicolumn{1}{l}{} & \multicolumn{1}{l}{} & \multicolumn{1}{l}{} &  & 0 & 0 & 0 & 0 & 0 &  &  \\ \cline{1-4} \cline{6-10}
\multicolumn{1}{|c|}{$\mathbf{A_{0}}$} & \multicolumn{1}{c|}{$\mathbf{B_{0}}$} & \multicolumn{1}{c|}{$\mathbf{C_{out}}$} & \multicolumn{1}{c|}{$\mathbf{S_{0}}$} &  & 0 & 0 & 1 & 0 & 1 &  &  \\ \cline{1-4} \cline{6-10}
\multicolumn{1}{|c|}{0} & \multicolumn{1}{c|}{0} & \multicolumn{1}{c|}{0} & \multicolumn{1}{c|}{0} &  & 0 & 1 & 0 & 1 & 1 &  &  \\ \cline{1-4} \cline{6-10}
\multicolumn{1}{|c|}{0} & \multicolumn{1}{c|}{1} & \multicolumn{1}{c|}{1} & \multicolumn{1}{c|}{0} &  & 0 & 1 & 1 & 1 & 1 &  &  \\ \cline{1-4} \cline{6-10}
\multicolumn{1}{|c|}{1} & \multicolumn{1}{c|}{0} & \multicolumn{1}{c|}{0} & \multicolumn{1}{c|}{1} &  & 1 & 0 & 0 & 0 & 1 &  &  \\ \cline{1-4} \cline{6-10}
\multicolumn{1}{|c|}{1} & \multicolumn{1}{c|}{1} & \multicolumn{1}{c|}{1} & \multicolumn{1}{c|}{1} &  & 1 & 0 & 1 & 1 & 0 &  &  \\ \cline{1-4} \cline{6-10}
\multicolumn{1}{l}{} & \multicolumn{1}{l}{} & \multicolumn{1}{l}{} & \multicolumn{1}{l}{} &  & 1 & 1 & 0 & 1 & 1 &  &  \\ \cline{6-10}
\multicolumn{1}{l}{} & \multicolumn{1}{l}{} & \multicolumn{1}{l}{} & \multicolumn{1}{l}{} &  & 1 & 1 & 1 & 1 & 0 &  &  \\ \cline{6-10}
\end{tabular}
}
\vspace{-10pt}
\end{table}
\begin{table}[!ht]
\centering
\caption{An example for MED calculation of 2-bit approximate adder}
\label{aaex}
\resizebox{0.48\textwidth}{!}{
\begin{tabular}{|c|c|c|c|c|c|c|c|c|}
\hline
                                                                                  &                                                                                        &                                                &                                                                                        &                                                                                                 &                                                                                                  &                                 &                                                                                       &                                                                                         \\
\multirow{-2}{*}{\textbf{\begin{tabular}[c]{@{}c@{}}Iteration\\ No.\end{tabular}}} & \multirow{-2}{*}{\textbf{\begin{tabular}[c]{@{}c@{}}Loop \\ counter ($i$)\end{tabular}}} & \multirow{-2}{*}{\begin{tabular}[c]{@{}c@{}}\textbf{Input\ Matrix} \\ $\{\mathbf{C_{inE}}\mathbf{C_{inA}}\}$\end{tabular}} & \multirow{-2}{*}{\textbf{\begin{tabular}[c]{@{}c@{}}Inputs\\ $A_i$ $B_i$\end{tabular}}} & \multirow{-2}{*}{\textbf{\begin{tabular}[c]{@{}c@{}}Exact\\ $C_{outE}$  $Sum_{E}$\end{tabular}}} & \multirow{-2}{*}{\textbf{\begin{tabular}[c]{@{}c@{}}Approx\\ $C_{outA}$  $Sum_{A}$\end{tabular}}} & \multirow{-2}{*}{\textbf{diff}} & \multirow{-2}{*}{\textbf{\begin{tabular}[c]{@{}c@{}}Matrix\\ Operation\end{tabular}}} & \multirow{-2}{*}{\textbf{\begin{tabular}[c]{@{}c@{}}Unchanged\\ Matrices\end{tabular}}} \\ \hline
1 & 0 & LMAT00 & 00 & {\color{blue}0}0  & {\color{blue}0}0 & 0 & HMAT{\color{blue} 00}=$\left|\begin{array}{cc} 0\textcolor{red}{+1} & 0  \\ 0 & 0 \end{array}\right|$ & \begin{tabular}[c]{@{}c@{}}HMAT01\\ HMAT10\\HMAT11\end{tabular} \\ \hline
2 & 0 & LMAT00 & 01 & {\color{blue}0}1 & {\color{blue}1}0 & 1 & HMAT{\color{blue} 01}=$\left|\begin{array}{cc} 0 & 0  \\ 0\textcolor{red}{+1} & 0 \end{array}\right|$ & \begin{tabular}[c]{@{}c@{}}HMAT00\\ HMAT10\\HMAT11\end{tabular} \\ \hline
3 & 0 & LMAT00 & 10 & {\color{blue}0}1 & {\color{blue}0}1 & 0 & HMAT{\color{blue} 00}=$\left|\begin{array}{cc} 1\textcolor{red}{+1} & 0  \\ 0 & 0 \end{array}\right|$ & \begin{tabular}[c]{@{}c@{}}HMAT01\\ HMAT10\\HMAT11\end{tabular} \\ \hline
4 & 0 & LMAT00 & 11 & {\color{blue}1}0 & {\color{blue}1}1 & -1 & HMAT{\color{blue} 11}=$\left|\begin{array}{cc} 0 & 0  \\ 0 & 0\textcolor{red}{+1} \end{array}\right|$  & \begin{tabular}[c]{@{}c@{}}HMAT00\\ HMAT01\\HMAT10\end{tabular} \\ \hline
5 & 1 & LMAT00 & 00 & {\color{blue}0}0 & {\color{blue}0}0 & 0 & HMAT{\color{blue} 00}=$\left|\begin{array}{cc} 0\textcolor{red}{+2} & 0  \\ 0\textcolor{red}{+0} & 0\textcolor{red}{+0} \\ 0 & 0 \\ 0 & 0 \end{array}\right|$  & \begin{tabular}[c]{@{}c@{}}HMAT01\\ HMAT10\\HMAT11\end{tabular} \\ \hline
6 & 1 & LMAT01 & 00 & {\color{blue}0}0 & {\color{blue}0}1 & -2 & HMAT{\color{blue} 00}=$\left|\begin{array}{cc} 2 & 0  \\ 0 & 0\textcolor{red}{+1} \\ 0 & 0\textcolor{red}{+0} \\ 0 & 0\textcolor{red}{+0} \end{array}\right|$ & \begin{tabular}[c]{@{}c@{}}HMAT01\\ HMAT10\\HMAT11\end{tabular}  \\ \hline
7 & 1 & LMAT10 & 00 & N/A & N/A & N/A & N/A & All matrices \\ \hline
8 & 1 & LMAT11 & 00 & {\color{blue}0}1 & {\color{blue}0}1 & 0 & HMAT{\color{blue} 00}=$\left|\begin{array}{cc} 2\textcolor{red}{+0} & 0  \\ 0\textcolor{red}{+0} & 1\textcolor{red}{+1} \\ 0 & 0 \\ 0 & 0 \end{array}\right|$  & \begin{tabular}[c]{@{}c@{}}HMAT01\\ HMAT10\\HMAT11\end{tabular} \\ \hline
9 & 1 & LMAT00 & 01 & {\color{blue}0}1 & {\color{blue}0}1 & 0 & HMAT{\color{blue} 00}=$\left|\begin{array}{cc} 2\textcolor{red}{+2} & 0  \\ 0\textcolor{red}{+0} & 2\textcolor{red}{+0} \\ 0 & 0 \\ 0 & 0 \end{array}\right|$  & \begin{tabular}[c]{@{}c@{}}HMAT01\\ HMAT10\\HMAT11\end{tabular} \\ \hline
10 & 1 & LMAT01 & 01 & {\color{blue}0}1 & {\color{blue}1}0 & 2 & HMAT{\color{blue} 01}=$\left|\begin{array}{cc} 0 & 0  \\ 0\textcolor{red}{+0} & 0 \\ 0\textcolor{red}{+0} & 0 \\ 0\textcolor{red}{+1} & 0 \end{array}\right|$  & \begin{tabular}[c]{@{}c@{}}HMAT00\\ HMAT10\\HMAT11\end{tabular} \\ \hline
11 & 1 & LMAT10 & 01 & N/A & N/A & N/A & N/A & All matrices \\ \hline
12 & 1 & LMAT11 & 01 & {\color{blue}1}0 & {\color{blue}1}0 & 0 & HMAT{\color{blue} 11}=$\left|\begin{array}{cc} 0\textcolor{red}{+0} & 0  \\ 0\textcolor{red}{+1} & 0\textcolor{red}{+0} \\ 0 & 0 \\ 0 & 0 \end{array}\right|$  & \begin{tabular}[c]{@{}c@{}}HMAT00\\ HMAT01\\HMAT10\end{tabular} \\ \hline
13 & 1 & LMAT00 & 10 & {\color{blue}0}1 & {\color{blue}1}1 & 0 & HMAT{\color{blue} 01}=$\left|\begin{array}{cc} 0\textcolor{red}{+2} & 0  \\ 0\textcolor{red}{+0} & 0\textcolor{red}{+0} \\ 0 & 0 \\ 1 & 0 \end{array}\right|$  & \begin{tabular}[c]{@{}c@{}}HMAT00\\ HMAT10\\HMAT11\end{tabular} \\ \hline
14 & 1 & LMAT01 & 10 & {\color{blue}0}1 & {\color{blue}1}1 & 0 & HMAT{\color{blue} 01}=$\left|\begin{array}{cc} 2\textcolor{red}{+0} & 0  \\ 0\textcolor{red}{+1} & 0\textcolor{red}{+0} \\ 0 & 0 \\ 1 & 0 \end{array}\right|$  & \begin{tabular}[c]{@{}c@{}}HMAT00\\ HMAT10\\HMAT11\end{tabular} \\ \hline
15 & 1 & LMAT10 & 10 & N/A & N/A & N/A & N/A & All matrices \\ \hline
16 & 1 & LMAT11 & 10 & {\color{blue}1}0 & {\color{blue}1}1 & -2 & HMAT{\color{blue} 11}=$\left|\begin{array}{cc} 0 & 0  \\ 0 & 1\textcolor{red}{+0} \\ 0 & 0\textcolor{red}{+0} \\ 0 & 0\textcolor{red}{+1} \end{array}\right|$  & \begin{tabular}[c]{@{}c@{}}HMAT00\\ HMAT01\\HMAT10\end{tabular} \\ \hline
17 & 1 & LMAT00 & 11 & {\color{blue}1}0 & {\color{blue}1}1 & -2 & HMAT{\color{blue} 11}=$\left|\begin{array}{cc} 0 & 0  \\ 0 & 1\textcolor{red}{+0} \\ 0 & 0\textcolor{red}{+2} \\ 0 & 1\textcolor{red}{+0} \end{array}\right|$  & \begin{tabular}[c]{@{}c@{}}HMAT00\\ HMAT01\\HMAT10\end{tabular} \\ \hline
18 & 1 & LMAT01 & 11 & {\color{blue}1}0 & {\color{blue}1}0 & 0 & HMAT{\color{blue} 11}=$\left|\begin{array}{cc} 0\textcolor{red}{+0} & 0  \\ 0\textcolor{red}{+1} & 1\textcolor{red}{+0} \\ 0 & 2 \\ 0 & 1 \end{array}\right|$  & \begin{tabular}[c]{@{}c@{}}HMAT00\\ HMAT01\\HMAT10\end{tabular} \\ \hline
19 & 1 & LMAT10 & 11 & N/A & N/A & N/A & N/A & All matrices \\ \hline
20 & 1 & LMAT11 & 11 & {\color{blue}1}1 & {\color{blue}1}0 & 2 & HMAT{\color{blue} 11}=$\left|\begin{array}{cc} 0\textcolor{red}{+0} & 0  \\ 1\textcolor{red}{+1} & 1\textcolor{red}{+0} \\ 0 & 2 \\ 0 & 1 \end{array}\right|$  & \begin{tabular}[c]{@{}c@{}}HMAT00\\ HMAT01\\HMAT10\end{tabular} \\ \hline
\end{tabular}}
\end{table}

 Four $2\times2$ matrices $HMAT00$, $HMAT01$, $HMAT10$, $HMAT11$0 are initialized to zero. The four matrix elements correspond to a difference of $+0,-0,+1,-1$ between exact and approximate adders. For $1^{st}$ iteration, ${00}$ input condition is considered for which carry-out for both accurate and approximate adder is 0. Hence, the matrix to be modified is $HMAT00$. The difference of the sum bit defined as $2^i(SUM_E-SUM_A)$ computes to 0 corresponding to the $HMAT00[0,0]$ index position. Element 1 in input matrix $HMAT00[0,0]$ index position is then copied to $HMAT00[0,0]$. Thus new $HMAT00$ becomes $\left | 1\ 0\ ;0\ 0 \right |_{2\times2}$ which means there is one input combination with zero difference while producing a $0$ carry-out bit in both exact and approximate adder.
All the other matrices remains unchanged. Similar operations are performed for next 3 possible input combinations. After $4$ iterations, one can observe that only $3$ matrices are modified which would then be used in subsequent iterations as input matrices. For $i=1$, four $4\times2$ new matrices $HMAT00$, $HMAT01$, $HMAT10$, $HMAT11$ are initialized to zero. Here, the $0^{th}$ column index positions refer to sum difference of $+0,+1,+2,+3$ while $1^{st}$ column index positions correspond to sum difference of $-0,-1,-2,-3$ between exact and approximate adder. Let us consider an intermediate iteration step 6. In this step, an input combination of ${00}$ is considered for MSB with exact carry-in of $0$ and approximate carry-in of $1$. Hence, $LMAT01$ is used as an input matrix. The carry-out bits for such a combination is $0$ for both exact and inexact additions. Hence, a matrix operation is performed on the $HMAT00$ matrix. The difference is computed as -2. The $[0,0]$, $[1,0]$, $[1,1]$ index position element of $LMAT01$ is added to $[1,2]$, $[1,1]$, $[1,3]$ index position element of $HMAT00$ respectively (refer Eqn. \ref{eq.im}). For instance, the $[1,0]$ index position of $LMAT00$ correspond to difference of $+1$. The new difference calculated after iteration $6$ would be $-2+1=-1$ which represents the  $[1,1]$ index position. Hence, the element $1$ in $[1,0]$ index of $LMAT01$ is added to $[1,1]$ index of $HMAT00$. All the remaining iterations follow the same principle. It should be noted that no operation is performed during iterations $7,11,15,19$ since those iterations correspond to an invalid exact and approximate carry-in combination of \{$1\ 0$\}.  
\section{Experimental Results and Analysis}
MED calculation using exhaustive simulation method have a time complexity of $\mathcal{O}(2^{2m})$. On the contrary, the asymptotic runtime of our MED computation method is $\mathcal{O}(m)$ for $k\ll m$. The total number of iterations can be represented using Eqn. \ref{eq.iter}. We can observe that as $k\to m$, the number of iterations becomes comparable to the exhaustive method. However, $k$ is typically $1$ for well-known LSB approximate adders such as LOA and AMA. For our experiments and analysis, we have considered $k=1$.
\small
    \begin{equation}
        \centering
        \label{eq.iter}
        \begin{split}
No.\ of\ iterations\ = \frac{m}{k} (2^{2k+2})
        \end{split}
    \end{equation}
    \begin{equation}
        \centering
        \label{eq.run}
        \begin{split}
Runtime\ Speed-up = \frac{MC\ sampling\ method\ runtime}{Proposed\ method\ runtime}
        \end{split}
    \end{equation}
\normalsize
The proposed algorithm is implemented in C++ to evaluate the MED parameter of approximate LSB adders. Several $16$-bit and $32$-bit approximate LSB adders are generated randomly whose MED value is then computed using our proposed method. We have compared the simulation runtime of our method with the Monte-Carlo (MC) sampling method presented in \cite{venkatesan2011macaco}. All the simulations are done in Linux operating system environment with Intel I5@3.2GHz processor core.

\begin{figure}[!ht]
    \centering
    \begin{subfigure}{0.2\textwidth}
        \centering
        \includegraphics[width=\textwidth]{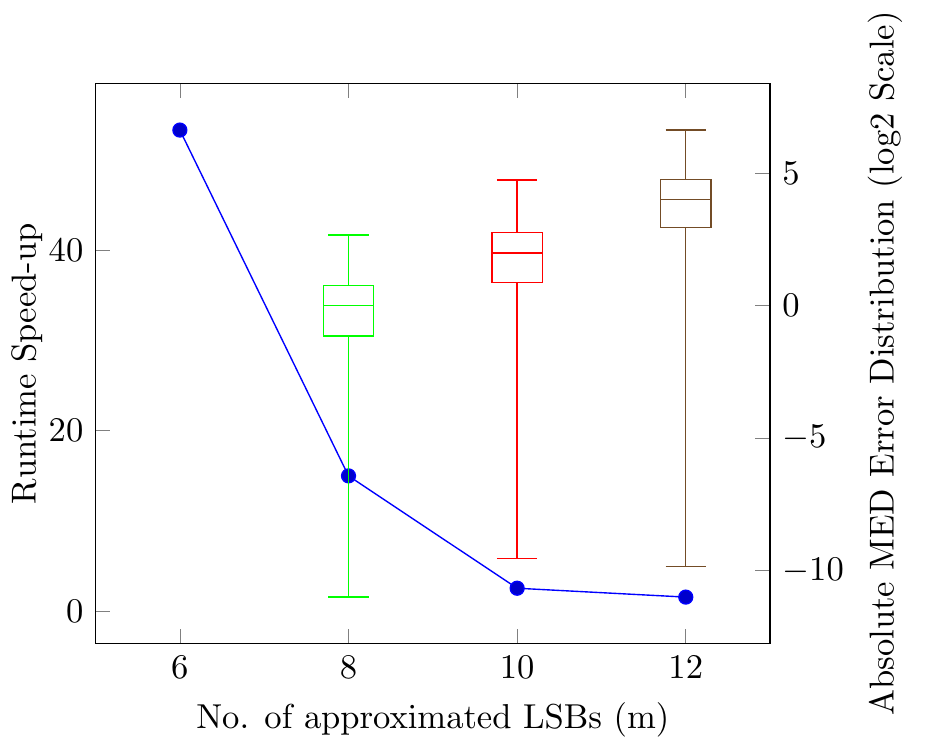}
        \caption{$n=16, k=1, S = 2^{12}$}
        \label{fig:1}
        \vspace{-0.2cm}
    \end{subfigure}%
    \begin{subfigure}[]{0.2\textwidth}
        \centering
        \includegraphics[width=\textwidth]{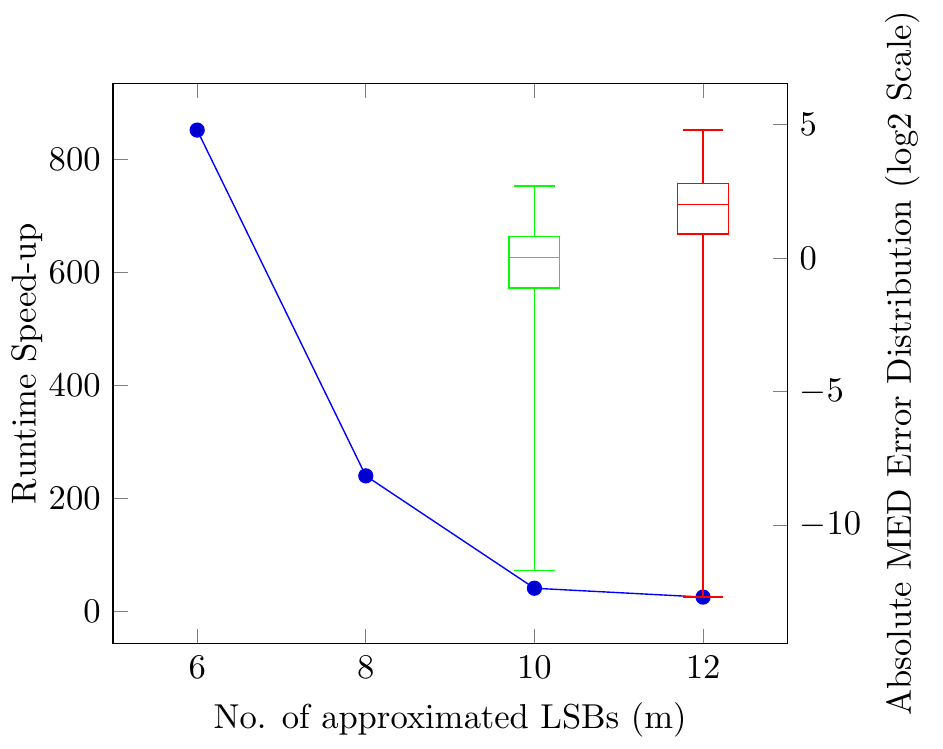}
        \caption{$n=16, k=1, S = 2^{16}$}
        \label{fig:2}
        \vspace{-0.2cm}
    \end{subfigure}
    \begin{subfigure}[]{0.2\textwidth}
        \centering
        \includegraphics[width=\textwidth]{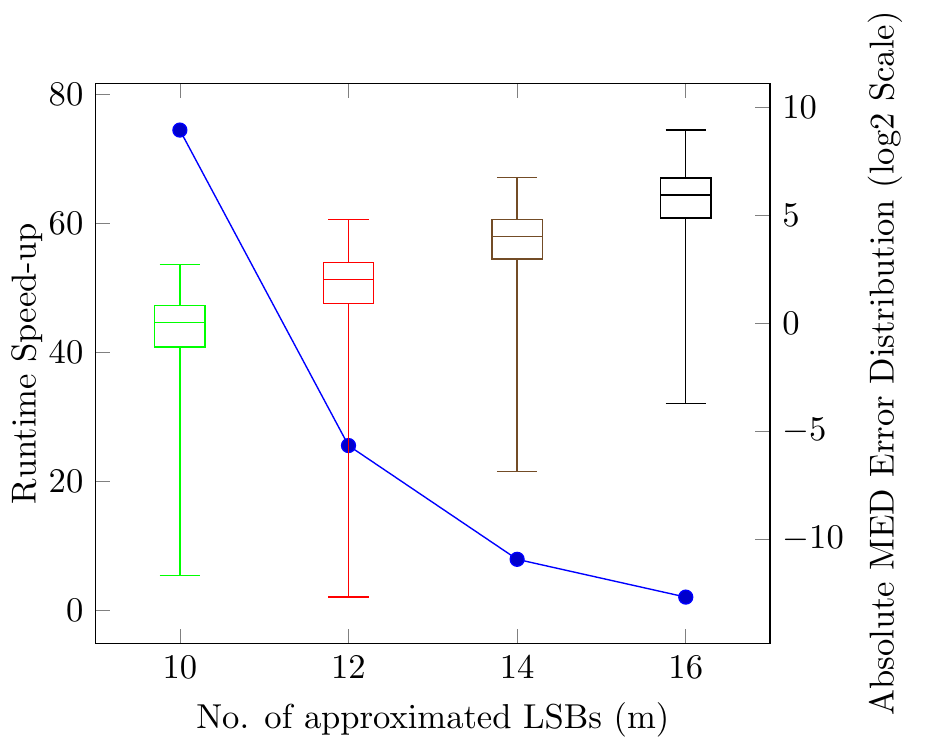}
        \caption{$n=32, k=1, S = 2^{16}$}
        \label{fig:3}
        \vspace{-0.2cm}
    \end{subfigure}%
    \begin{subfigure}[]{0.2\textwidth}
        \centering
        \includegraphics[width=\textwidth]{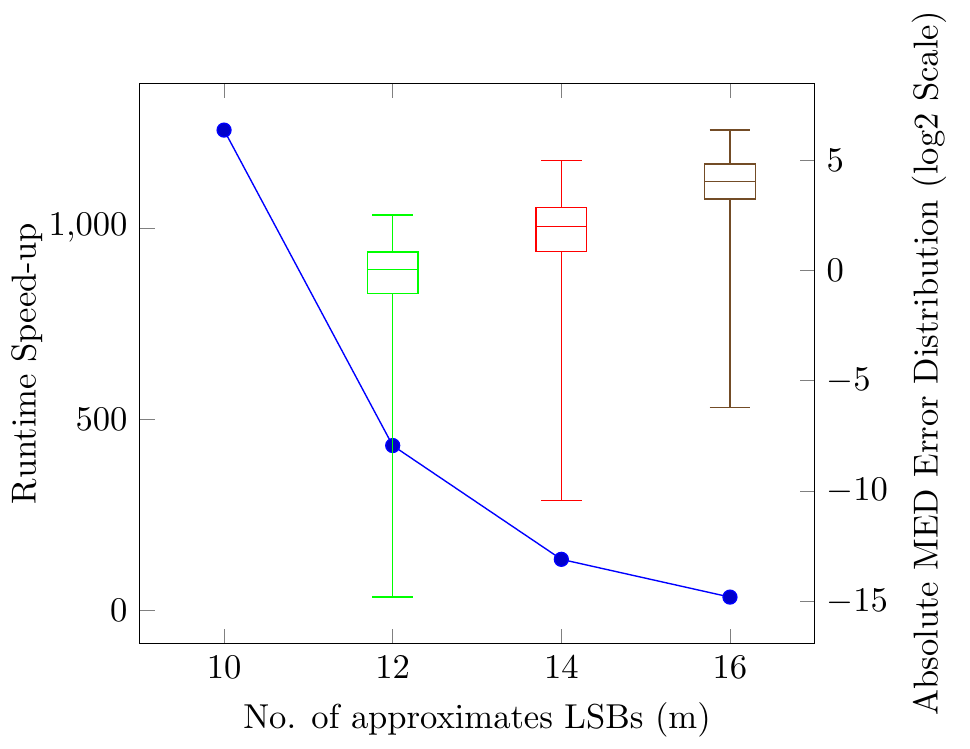}
        \caption{$n=32, k=1, S = 2^{20}$}
        \label{fig:4}
        \vspace{-0.2cm}
    \end{subfigure}
    \caption{Speed-up in runtime of proposed method as compared to MC sampling method}
    \label{fig:rc}
    \vspace{-0.3cm}
\end{figure}
Fig. \ref{fig:1} and Fig. \ref{fig:2} shows the speed-up in runtime (Eqn. \ref{eq.run}) of the proposed method compared to MC sampling method with samples(S) equal to $2^{12}$ and $2^{16}$ respectively for the $16$-bit adder. The results are plotted for different values of $m = 6,8,10,12$. Similarly, the results for $32$-bit adder with $m = 10,12,14,16$ are also represented in Fig. \ref{fig:3} and Fig. \ref{fig:4} for S equal to $2^{16}$ and $2^{20}$ respectively. Since the MC sampling method only provides an estimate of the actual MED value, there would be some finite error in the calculated MED. The error distribution of calculated MED using MC sampling method in log2 scale is also plotted in the form of box plots. We have considered random $5000$ approximate LSB adders for each case to plot the error distribution. The maximum, minimum, median, first and the third quartile of the MED error distribution is shown for MC sampling method. It can be observed that the proposed method provides higher speed-up for all values of $m$ considered. Since the number of iterations in $Med\_Cal$ increases with $m$; there is also a decrease in the speed. However, at the same time as $m$ increases, the error in the MC sampling method also increases as shown by the plots. Experimental results show that our proposed method is approximately $25$ times faster than MC sampling method with $2^{16}$ samples for $n=16\ and\  m=12$. Compared to our accurate technique, MC sampling method have an error median of $4$ with maximum absolute error as $23$ for $5000$ approximate LSB adder cases. Similarly for $n=32\ and\  m=16$, the proposed method is roughly $2$ and $34$ times faster compared to MC sampling method with $2^{16}$ and $2^{20}$ samples respectively. The respective median error for MC sampling method is $35$ and $16.21$, whereas the maximum error observed was $80$ and $45$ respectively. The number of exhaustive simulations required for $m$ equal to $6$, $8$, and $10$ is $2^{12}$, $2^{16}$, and $2^{20}$ respectively. Hence, there is no error for respective sample size which is shown by the absence of MED error distribution plot in Fig. \ref{fig:1}, \ref{fig:2}, and \ref{fig:3}.
\section{Conclusions}
This article has proposed a new efficient algorithm to determine accurate MED of approximate LSB adders. The execution time of the proposed method has a linear dependence on the number of LSBs approximated, thus making it much faster than the exhaustive technique. Experimental analysis shows that for $k$ taken as unity, the proposed method is superior to MC sampling method. The developed MED evaluation technique can be modified to compute other errors metrics such as ER, MSED, MRED
    for approximate LSB adders. In future, we wish to extend our work by developing algorithms which would compute and analyze all error metrics related to approximate LSB adders.

\bibliographystyle{IEEEtran}

\end{document}